\documentclass[aps,prb,twocolumn,amsmath,amssymb]{revtex4-2}
\usepackage{graphicx}
\usepackage{bm}
\usepackage[dvipdfmx]{color}
\usepackage[dvipdfmx,colorlinks=true,citecolor=blue,linkcolor=blue]{hyperref}
\usepackage{xcolor}
\allowdisplaybreaks[4]

\begin{document}
\title{
    Superconductivity and the quasiparticle mass enhancement near the CDW critical point using Bethe-Salpeter method: Application to cuprates
}

\author{Youichi Yamakawa}
\email[]{yamakawa.youichi.p1@f.mail.nagoya-u.ac.jp}
\affiliation{Department of Physics, Nagoya University, Nagoya 464-8602, Nagoya 464-8602, Japan}

\author{Hiroshi Kontani}
\affiliation{Department of Physics, Nagoya University, Nagoya 464-8602, Nagoya 464-8602, Japan}

\date{\today}

\begin{abstract}
    In recent years, charge-channel order in strongly correlated metals has attracted much attention. 
    Representative examples include electronic nematic order in cuprates and iron-based superconductors, and Star-of-David order in kagome metals. 
    Critical phenomena and unconventional superconductivity arising from fluctuations of such charge-channel orders are central issues today; however, the essential role is played by many-body effects (vertex corrections) beyond the mean-field approximation, and their origin and computational methods have not been established. 
    In this study, we propose the Bethe-Salpeter equation method to evaluate electron-electron interactions in two-dimensional Hubbard models beyond the mean-field approximation. 
    Based on the Baym-Kadanoff conserving approximation, we find that an attractive interaction in the charge channel emerges from the Aslamazov-Larkin vertex corrections that describe the interference processes among spin fluctuations. 
    Applying this method to the square-lattice Hubbard model shows that the cooperation of attractive charge fluctuations and repulsive spin fluctuations yields high-$T_c$ $d$-wave superconductivity together with enhanced effective mass. 
    These results provide a natural explanation for the phase diagram of cuprate superconductors, in which $d$-wave superconductivity is strongly enhanced near the charge-order critical point.
    The theory can also be applied to iron-based and nickelate superconductors, suggesting broad potential for future applications.
\end{abstract}


\sloppy

\maketitle

\section{Introduction}\label{sec1}
    In recent years, many interesting quantum phases involving charge and orbital degrees of freedom have been discovered in strongly correlated metals. 
    A representative example is the charge-density-wave (CDW) order in cuprate high-$T_c$ superconductors, which has been observed by $X$-ray scattering~\cite{Ghiringhelli2012_LR_CDW_YNdYBCO, Comin2014_Xray-STM_CDW, Tabis2014_Xray_CDW_Hg1201,Chang2016_CDW_YBCO_mag, Gerber2015_YBCO_3D_CDW, Kim2021_YBCO_CDW_uniaxial, Neto2015_Xray_CDW, Peng2018_Xray_CDW, Arpaia2019_dynamicalCDW}, 
    scanning tunneling microscopy (STM)~\cite{Kohsaka2012_STM_CDW, Fujita2014_STM_CDW}, 
    and nuclear magnetic resonance (NMR)~\cite{Wu2015_NMR_CDW, Kawasaki2017_NMR_CDW, Vinograd2021_NMR_CDW}. 
    In addition, the electronic nematic order in iron-based superconductors~\cite{Fisher2010_nematic, Bohmer2018_nematic_review}, the $2\times 2$ CDW in kagome metals~\cite{Ortiz2020_CsV3Sb5_PRL, Li2022_KV3Sb5_2x2CDW}, and loop-current order~\cite{Jiang2021_KV3Sb5_LC, Xing2024_RbV3Sb5_LC, Deng2024_AV3Sb5_ChiralSC} have attracted considerable attention. 
    These charge-channel orders arising from strong electron correlations do not necessarily accompany conventional spin-density-wave (SDW) order. 
    In strongly correlated electron systems, the local Coulomb repulsion $U$ is comparable to the bandwidth and promotes spin polarization, so conventional mean-field-level approximations predict the pure SDW order~\cite{Moriya1985,Mahan2000}. 
    Therefore, to understand the charge-channel orders without magnetization, 
    it is essential to construct many-electron theories beyond the mean-field approximation.
    This fact strongly motivates researchers to improve previous theories of electronic correlations~\cite{Keimer2015_Review, Chang2016_CDW_YBCO_mag}.

    Charge-channel orders in the above-mentioned strongly correlated metals are considered to be closely related to unconventional superconductivity. 
    In $d$-wave cuprate superconductors, such as YBCO, both the superconducting transition temperature $T_c$ and the upper critical field $H_{c2}$ take their maximum values at the quantum critical point (QCP) at the end of the CDW phase for many cuprate compounds~\cite{Ramshaw2015_mass, Chang2012_Hc2_LSCO, Tabis2014_Xray_CDW_Hg1201}. 
    That is, superconductivity is strongly enhanced near the QCP of charge-channel order. 
    Similar enhancement of superconductivity is also observed at the end of the nematic order in the $s$-wave iron-based superconductor Fe(Se,Te)~\cite{Mukasa2023FeSeTe} and at the end of the $2\times 2$ CDW in the kagome-lattice superconductor CsV$_3$Sb$_5$~\cite{Chen2021CsV3Sb5}. 
    Understanding unconventional superconductivity in these systems requires theoretical methods based on the Hubbard model that can correctly describe not only spin fluctuations but also charge-channel orders and their critical fluctuations.

    Research on $d$-wave superconductivity and spin/charge stripe orders has advanced through cluster dynamical mean-field theory (cDMFT)~\cite{Maier2005_cDMFT, Schafer2021_cDMFT}, variational Monte Carlo (VMC)~\cite{Yokoyama1987_VMC, Sandro2001_VMC, Karakuzu2022_VMC}, and density-matrix renormalization group (DMRG)~\cite{White1992_DMRG, Leblanc2015_DMRG, Stoudenmire2012_DMRG}. 
    These theoretical methods can be applied even in the strong-coupling regime. 
    While VMC and DMRG are typically performed on finite-size systems, the main limitation to cDMFT is related to the treatment of short-range correlations only.
    In addition, analyses based on field-theoretical approaches such as the functional renormalization group \cite{Salmhofer2001fRG, Metzner2012fRG} and the parquet renormalization group \cite{Zheleznyak1997_pRG} have revealed that various charge-channel orders emerge due to many-body effects beyond the mean-field approximation~\cite{Tazai2022_kagome, Tazai2016_fRG_orbital, Tazai2021_fRG_loop, Tsuchiizu2013fRG, Tsuchiizu2018cuprate, Chubukov2016_pRG_FeSC}. 

    The authors have focused on the Aslamazov-Larkin (AL) type vertex correction, which describes the coupling between charge and spin fluctuations, and have explained the orbital order and nematic order in iron-based superconductors~\cite{Onari2012_SCVC, Yamakawa2016FeSe}, the CDW in cuprates~\cite{Yamakawa2015cuprate, Kawaguchi2017DW, Tsuchiizu2018cuprate}, and the $2\times 2$ CDW in kagome metals~\cite{Tazai2022_kagome}. 
    The AL-type vertex correction describes magnon-pair formation in metals ($\langle s_i s_j \rangle \ne 0$), under which a variety of charge-channel orders appear without spin order (i.e., $\langle s_i \rangle = 0$)~\cite{Kontani2021_review}. 
    This theory satisfies the criteria of the 
    Baym-Kadanoff conserving approximation~\cite{BaymKadanoff1961, Tazai2023_Rigorous}, and it is applicable to various multi-orbital Hubbard models. 
    The charge-channel orders produced by this mechanism correspond to modulations $\delta t_{i,j}$ of the hopping integral, and are hereafter referred to as bond order (BO).

    At the quantum critical point (QCP) at the end of an ordered phase in metals, well-developed quantum fluctuations of bosonic modes can give rise to unconventional superconductivity and quantum critical phenomena. 
    The most extensively studied example is the QCP of the spin-density-wave (SDW) phase~\cite{Hertz1976_QCP, Millis1993_QCP}.
    The methods for calculating the interaction $V_{\rm spin}$ mediated by developed spin fluctuations, as well as the superconductivity mediated by $V_{\rm spin}$, are now almost established~\cite{Bickers1989_FLEX_SC, Bickers1991_FLEX}. 
    In contrast, the theory for the QCP of the bond-order phase is still incomplete. 
    The calculation methods for the interaction $V_{\mathrm{BO}}$ mediated by developed bond-order fluctuations have remained at a qualitative level, involving phenomenological parameters. If a method for calculating $V_{\mathrm{BO}}$ with high accuracy is established, it will become possible to quantitatively study unconventional superconductivity and quantum critical phenomena mediated by bond-order fluctuations. Therefore, this theoretical problem is of great importance.

    In this study, we propose the ``Bethe-Salpeter (BS) equation method" to quantitatively determine $V_{\mathrm{BO}}$ beyond the mean-field approximation. This theory is based on the Baym-Kadanoff conserving approximation and systematically incorporates vertex corrections, such as the Maki-Thompson term and the Aslamazov-Larkin term.
    These vertex corrections describe the coupling among spin fluctuations.
    The $V_{\mathrm{BO}}$ obtained in this theory exhibits a pronounced momentum dependence and increases divergently near the QCP. 
    We then apply the theory to the single-orbital square-lattice Hubbard model, which is an effective model for cuprate high-$T_c$ superconductors. 
    As a result, we find that near the QCP of the $d$-wave bond order ($p \sim 0.2$), spin fluctuations and bond-order fluctuations cooperate to strongly enhance $d$-wave superconductivity (high $T_c$ and $H_{c2}$) and give rise to quantum critical phenomena (large $m^*$)  consistent with experimental observations.
    This theory provides a clear physical understanding of the characteristic electronic state near the QCP in cuprate high-$T_c$ superconductors.
    In addition, the present theoretical framework can be easily applied to more complex multi-orbital metals, such as iron-based superconductors and Ni-based oxide superconductors. 

    The remainder of this paper is organized as follows.
    In Sec.~\ref{sec2}, we introduce the model and the FLEX approximation.
    In Sec.~\ref{sec3}, we present the Bethe-Salpeter (BS) formalism.
    In Sec.~\ref{sec4}, we discuss the mass enhancement.
    In Sec.~\ref{sec5}, we analyze superconductivity.
    In Sec.~\ref{sec6}, we present the phase diagram.
    Finally, Sec.~\ref{sec7} is devoted to discussion and conclusions.

\section{Model and FLEX}\label{sec2}
    \begin{figure}[!htb]
        \includegraphics[width=1\linewidth]{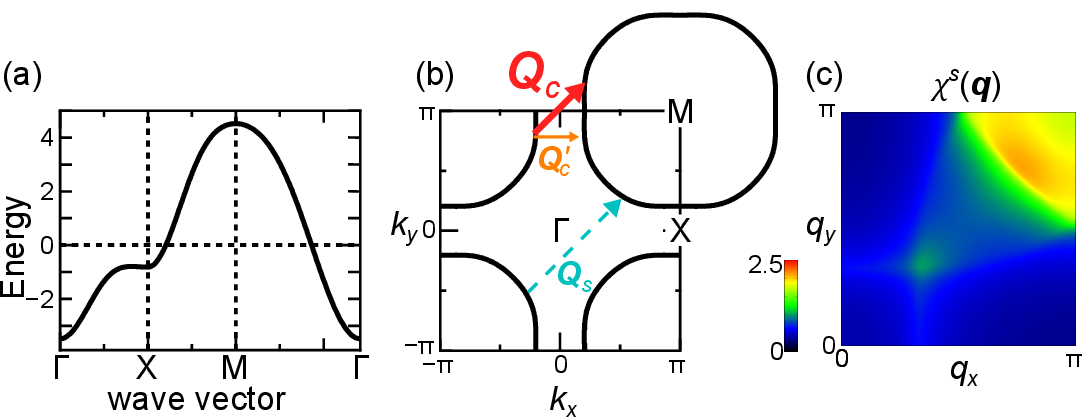}
        \caption{
            (a) Band structure and (b) Fermi surface for hole doping $p = 0.2$. 
            The nesting vector for spin fluctuations is $\bm{Q}_s \approx (\pi, \pi)$. 
            $\bm{Q}_c = (\delta, \delta)$ and $\bm{Q}_{c}' = (\delta, 0)$ are the nesting vectors associated with charge-channel fluctuations; the experimentally observed CDW corresponds to $\bm{Q}_{c}'$. 
            (c) Spin susceptibility $\chi^s(\bm{q})$ calculated within FLEX for $p = 0.2$, $U = 4.29$, and $T = 0.01$.
        }\label{fig1}
    \end{figure}
    First, we introduce the single-orbital Hubbard model and show the basic results of the fluctuation-exchange (FLEX) approximation, which will serve as the starting point for the BS equation discussed later. 
    The Hamiltonian is  
    \begin{align}
        H = \sum_{\bm{k}, \sigma} \epsilon_{\bm{k}} c_{\bm{k}, \sigma}^{\dagger} c_{\bm{k}, \sigma}
        + U \sum_{i} n_{i, \uparrow} n_{i, \downarrow},
    \end{align}
    where $c_{\bm{k}, \sigma}^{\dagger}$ creates an electron with wave vector $\bm{k}$ and spin $\sigma$, and $n_{i, \sigma} = c_{i, \sigma}^{\dagger} c_{i, \sigma}$ is the density operator. 
    $U$ is the local Coulomb interaction. 
    The band dispersion is 
    $\epsilon_{\bm{k}} = - 2t \left( \cos k_x + \cos k_y \right) - 4 t' \cos k_x \cos k_y - 2 t'' \left( \cos 2k_x + \cos 2k_y \right)$,
    and we use $t = 1$, $t' = -t/5$, and $t'' = t/6$ in this study.

    Fermi surface (FS) and band structure for hole doping $p = 0.2$ are shown in Figs~\ref{fig1}(a) and \ref{fig1}(b), respectively. 
    The nesting vector of spin fluctuations is $\bm{Q}_s \approx (\pi, \pi)$, while candidate wave vectors for charge-channel fluctuations are $\bm{Q}_c = (\delta, \delta)$ and $\bm{Q}_{c}' = (\delta, 0)$. The experimentally observed CDW corresponds to $\bm{Q}_{c}'$. 
    Unless otherwise noted, we use the nearest-neighbor hopping $|t| = 1$
    (corresponding to $\sim 0.5$~eV based on band-structure calculations~\cite{Pavarini2001_LDA,Elfimov2008_LDA})
    as the unit of energy.

    Figure~\ref{fig1}(c) presents the spin susceptibility calculated using FLEX,
    $\chi^s (\bm{q}) = \frac{\chi^0 (\bm{q})}{1 - \alpha_s (\bm{q})}$,
    where $\chi^0(\bm{q})$ is the irreducible susceptibility and $\alpha_s(\bm{q}) \equiv U \chi^0(\bm{q})$ is the $\bm{q}$-dependent spin Stoner factor. The spin Stoner factor $\alpha_s \equiv \max_{\bm{q}} \left[ \alpha_s(\bm{q}) \right]$ reaches unity at the SDW transition. For $U=4.29$, $T=0.01$, and $p=0.2$, we obtain $\alpha_s = 0.9$, indicating well-developed spin fluctuations around $\bm{Q}_s$.
    The present parameters are in the weak- to intermediate-coupling regime, where a FLEX-based approach is expected to be applicable. 
    The estimated values of $U/t$ for cuprates vary depending on the material and method, and values in the intermediate-coupling regime have also been reported~\cite{Jang2016_cRPA}. 

    In contrast, the charge susceptibility from FLEX,
    $\chi^c (\bm{q}) = \frac{\chi^0 (\bm{q})}{1 + \alpha_c (\bm{q})}, \quad \alpha_c(\bm{q}) \equiv -U \chi^0(\bm{q})$,
    is strongly suppressed by $U$. 
    This means that FLEX, a conserving approximation widely used for describing spin fluctuations and satisfying the Mermin-Wagner theorem~\cite{Bickers1989_FLEX,Bickers1989_FLEX_SC,Bickers1991_FLEX,Graser2010_FLEX_Fe,Sakakibara2024_FLEX_Ni}, 
    cannot reproduce charge-channel orders driven by local Coulomb repulsion.

\section{Bethe-Salpeter (BS) equation}\label{sec3}
    In this section, we introduce the Bethe-Salpeter (BS) equation formalism with full vertex corrections, which constitutes the central theoretical framework of this study. 
    \subsection{Formulation of the BS equation}
        \begin{figure}[!htb]
            \includegraphics[width=1\linewidth]{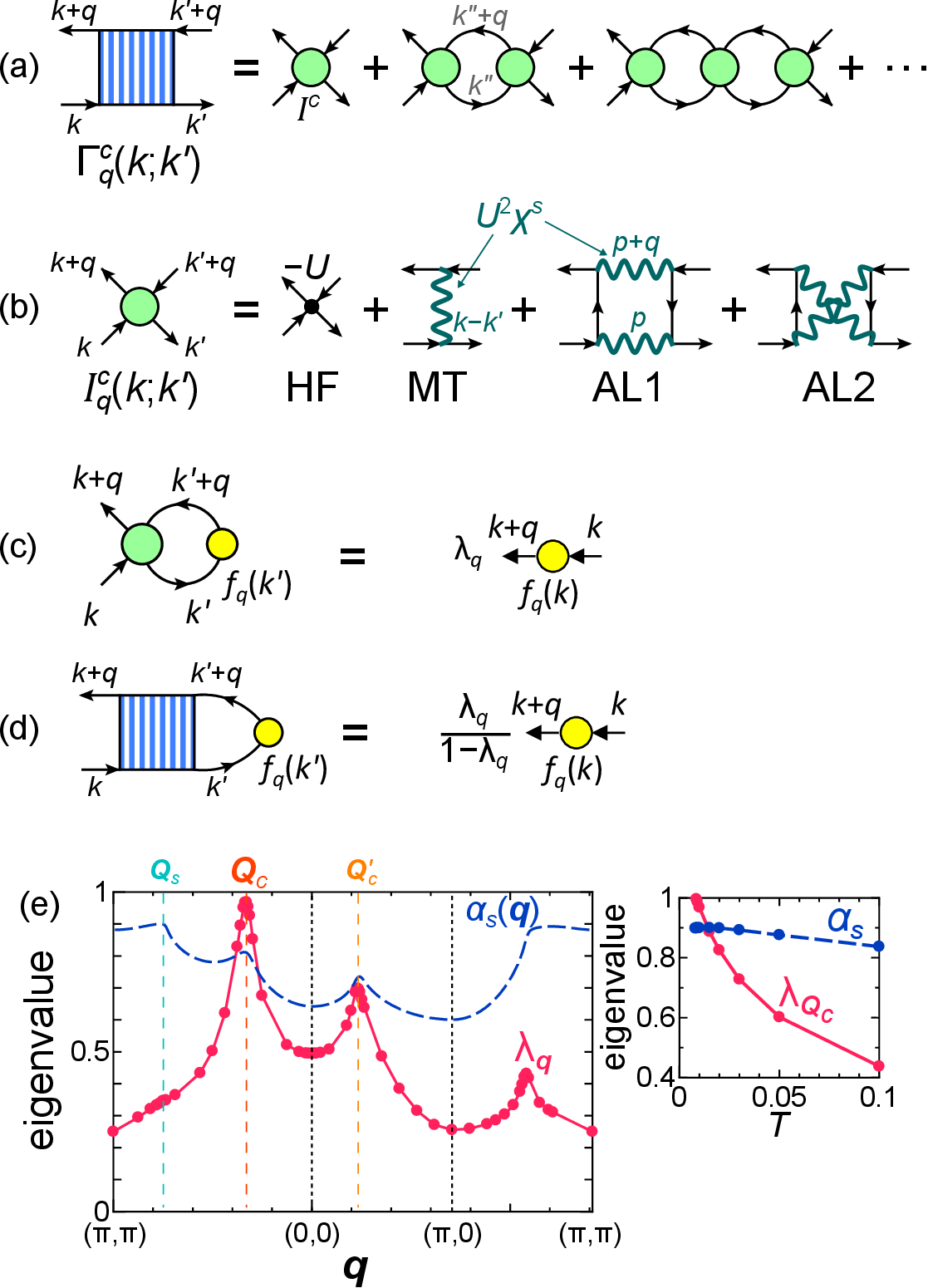}
            \caption{
                (a) Diagrammatic representation of the 
                BS equation for the full vertex $\Gamma^c_q (k; k')$. 
                $\Gamma_{\bm{q}}$ increases as $\propto I_{\bm{q}} / (1 - \lambda_{\bm{q}})$ near the eigenvalue $\lambda_{\bm{q}} = 1$, where $\lambda_{\bm{q}}$ corresponds to the instability in the charge channel, and $1/(1 - \lambda_{\bm{q}})$ to the charge susceptibility. 
                (b) Irreducible interaction $I^c$, consisting of one Hartree term, one 
                MT term (first-order in fluctuations), and two 
                AL terms (second-order in fluctuations). 
                (c) 
                DW equation with the form factor $f_{\bm{q}}(k)$ and eigenvalue $\lambda_{\bm{q}}$; the transition occurs at $\lambda_{\bm{q}} = 1$. 
                (d) Diagonalization equation for $\Gamma$, whose eigenvalue is $\lambda_{\bm{q}}/(1 - \lambda_{\bm{q}})$, diverging at $\lambda_{\bm{q}} \to 1$ (particle-hole ordering transition). 
                (e) $\bm{q}$-dependence of $\lambda_{\bm{q}}$ for $p = 0.2$, $U = 4.29$, and $T = 0.01$. 
                $\lambda_{\bm{q}}$ becomes large around $\bm{Q}_c = (\delta, \delta)$ and $\bm{Q}_{c}' = (\delta, 0)$.
                Inset: Temperature dependence of $\lambda_{\bm{Q}_c}$ and $\alpha_s$, showing that $\lambda_{\bm{Q}_c}$ surpasses $\alpha_s$ at low $T$, and the charge susceptibility diverges at $T_{\mathrm{CDW}} = 0.009$.
            }\label{fig2}
        \end{figure}

        By including the vertex corrections neglected in the FLEX~\cite{Bickers1989_FLEX_SC}, 
        the Bethe-Salpeter (BS) equation provides a framework to describe the development of charge-channel fluctuations beyond mean-field theory. 
        Within the microscopic Fermi-liquid theory~\cite{Abrikosov1963},
        the BS equation for the full electron-electron vertex $\Gamma^x_q(k,k')$ is given by
        \begin{widetext} 
        \begin{equation}
            \Gamma^x_q(k,k') = I^x_q(k,k') - \frac{T}{N} \sum_{k''} I^x_q(k,k'') G(k''+q) G(k'') \Gamma^x_q(k'',k'), 
            \label{eq-gamma0}
        \end{equation}
        \end{widetext}
        where $x = c$ denotes the charge channel and $x = s$ the spin channel. 
        Here, $k = (\bm{k}, i\epsilon_n)$ and $q = (\bm{q}, i\omega_l)$ represent the combined notations for wave vector and Matsubara frequency for fermions $\epsilon_n = (2n+1)\pi T$ and bosons $\omega_l = 2l\pi T$. 
        In the present study, we employ the Green's function obtained within the FLEX approximation, which is determined from the Dyson equation
        $ G(k) = 1/\left( i\epsilon_n - \varepsilon_{\bm{k}} + \mu - \Sigma_{\mathrm{FLEX}} (k) \right)$, 
        where $\Sigma_\mathrm{FLEX} (k)$ is the FLEX self-energy.
        The chemical potential $\mu$ is adjusted at each step of the self-consistent calculation so as to keep the filling.
        Solving Eq.~\eqref{eq-gamma0} self-consistently yields the full charge-channel vertex $\Gamma^c_q(k,k')$ [Fig.~\ref{fig2}(a)],
        where the internal momentum and Matsubara-frequency summations are fully taken into account.

        The vertex corrections are included in the irreducible interaction $I^c_q(k,k')$ in the particle-hole (p-h) channel [Fig.~\ref{fig2}(b)], which is derived from the one-loop approximation of the Luttinger-Ward functional $\Phi^{\rm LW}$~\cite{Tazai2023_Rigorous}. 
        The irreducible interaction $I$ is derived as $I = \delta^2 \Phi^\mathrm{LW} [G]/ \delta G^2 = \delta \Sigma_\mathrm{FLEX} [G] / \delta G$, ensuring that the self-energy and the BS kernel are constructed consistently within the Baym-Kadanoff conserving approximation~\cite{BaymKadanoff1961, Tazai2023_Rigorous}.
        It consists of the Hartree term by $U$ (or, more generally, Hartree-Fock (HF) term),
        the Maki-Thompson (MT) term, and two Aslamazov-Larkin (AL) terms.
        The explicit expressions for $I^c$ and $I^s$ are given in Appendix~\ref{app-BS}.

        If only the Hartree term is retained, 
        solving the BS equation~\eqref{eq-gamma0} reproduces the FLEX interaction $V^x = U^x + U^x \chi^x U^x$. 
        The MT term in $I^c$ [Fig.~\ref{fig2}(b)] corresponds to the spin susceptibility $\chi^s$ in FLEX, while $I^c$ also includes AL terms representing convolutions of spin susceptibilities. 
        Thus, the MT and AL terms represent vertex corrections beyond FLEX. 
        Solving Eq.~\eqref{eq-gamma0} with these terms generates higher-order diagrams, as shown in Fig.~\ref{fig7}(b) in Appendix \ref{app-BS}.
            
        In particular, the AL terms drive the CDW instability via quantum interference between SDW fluctuations, and successfully explain the nonmagnetic nematic order in FeSe~\cite{Yamakawa2016FeSe} and the CDW in CsV$_3$Sb$_5$~\cite{Tazai2022_kagome}.
        The MT term is also known to play an important role in current vertex corrections in cuprates~\cite{Kontani1999_Hall,Simard2021_CVC}.
        However, previous studies showed that the MT term alone does not fully account for the experimentally observed CDW tendencies in cuprates~\cite{Metlitski2010_CDW,Sachdev2013_CDW}. 
        The importance of the MT and AL terms has also been confirmed using unbiased many-body methods such as the functional renormalization group~\cite{Tazai2016_fRG_orbital, Tazai2021_fRG_loop, Tsuchiizu2018cuprate,Tsuchiizu2013fRG}.

        The superconducting (SC) gap equation is equivalent to the BS equation in the particle-particle (p-p) channel, where it is well established that the MT term drives unconventional SC states with sign-reversals~\cite{Moriya2000_highTc}. 
        Recently, the importance of the MT and AL terms in the particle-hole channel BS equation has been discovered as explained in Ref.~\cite{Kontani2021_review}.
        To elucidate their impact, we next analyze $\Gamma^c_q(k,k')$ using the density-wave (DW) equation.

    \subsection{Physical interpretation of $\Gamma_q(k,k')$}
        To analyze the structure of the full vertex $\Gamma^c_q(k,k')$ obtained from the Bethe-Salpeter equation, we express it in terms of the leading instability of the density-wave (DW) equation shown in Fig.~\ref{fig2}(c)~\cite{Tazai2023_Rigorous}.
        The DW equation is
        \begin{equation}
            \lambda_q f_q (k) = \frac{T}{N} \sum_{k'} I^c_q (k,k') G(k'+q) G(k') f_q (k').
            \label{eq-DW}
        \end{equation}
        Here, $\lambda_q$ is the largest eigenvalue of the DW equation, which characterizes the instability; the DW transition occurs at $\lambda_q = 1$. The corresponding eigenfunction $f_q(k)$ describes the nonlocal structure of the DW order form factor.

        Replacing the kernel $I^c$ in the DW equation with the full vertex $\Gamma^c$ yields eigenvalues of $f_q(k)$ as $\lambda_q/(1 - \lambda_q)$ [Fig.~\ref{fig2}(d)]. 
        Thus, for $\lambda_q \approx 1$, $\Gamma^c_{\bm{q}}$ is well approximated by
        \begin{align}
            \Gamma^c_{\bm{q}} (k,k') 
            \approx
            f_{\bm{q}}(k) \frac{\bar{I}^c_{\bm{q}}}{1 - \lambda_{\bm{q}}} f^*_{\bm{q}}(k'),
            \label{eq-gamma-approx}
        \end{align}
        which increases critically near $\bm{q} \approx \bm{Q}_c$.
        (Mathematically, Eq.~\eqref{eq-gamma-approx} corresponds to the maximum eigenvalue component in a singular value decomposition.)
    
        The proportional relation $\Gamma^c_{\bm{q}} \propto \bar{I}^c_{\bm{q}} / (1 - \lambda_{\bm{q}}) \propto \lambda_{\bm{q}} / (1 - \lambda_{\bm{q}})$ in Eq.~\eqref{eq-gamma-approx} holds exactly only when $\Gamma^c$ is obtained by solving the BS equation~\eqref{eq-gamma0} with full summations over all internal $\bm{k}$ and $\epsilon_n$. 
        We note that Eq. \eqref{eq-gamma-approx} is introduced only for physical interpretation and is not used in the numerical calculations.

        Once $\Gamma_q(k,k')$ is accurately determined, various physical quantities, such as the superconducting gap function and the self-energy, can be calculated.  
        While Eq.~\eqref{eq-gamma-approx} is quantitatively reliable only near the QCP, 
        in this study we solve the BS equation numerically to obtain $\Gamma^c$ with high precision for $q$, $k$, and $k'$, irrespective of the distance from the QCP.  
        (The $\Gamma^c$ obtained from the BS equation~\eqref{eq-gamma0} naturally satisfies Eq.~\eqref{eq-gamma-approx} near the QCP.)  
        This ``BS-eq. method" thus enables quantitative calculations of critical phenomena across a wide range of physical quantities.

        Figure~\ref{fig2}(e) shows the $\bm{q}$-dependence of $\lambda_{\bm{q}}$ for $U=4.29$, $T=0.01$, and $p=0.2$. 
        $\lambda_{\bm{q}}$ exhibits peaks at $\bm{Q}_c$ and $\bm{Q}_{c}'$, indicating the stabilization of charge-channel orders, 
        namely $d$-wave BO in cuprates,
        with these wave vectors~\cite{Kawaguchi2017DW,Tsuchiizu2018cuprate}.
        $\lambda_q$ and the MT-term contribution in the $q_x$-$q_y$ plane are shown in Appendix \ref{app-DW} (Fig.~\ref{fig9}).
        In the AL terms, nesting plays a more important role than in the Lindhard function because vertex corrections become large at low energies. Therefore, $\bm{Q}_c$ and $\bm{Q}_c'$ are determined by nesting between nearly parallel segments of the Fermi surface~\cite{Yamakawa2015cuprate}.
        The inset plots the temperature dependence of $\lambda_{\bm{Q}_c}$ and the spin Stoner factor $\alpha_s$. 
        As the temperature decreases, $\lambda_{\bm{Q}_c}$ overtakes $\alpha_s$, and the charge susceptibility $\propto 1/(1 - \lambda_{\bm{Q}_c})$ increases divergently toward the critical temperature $T_{\mathrm{CDW}} = 0.009$.

        The BO fluctuations originate from the Aslamazov-Larkin (AL) terms in the charge-channel irreducible vertex $I^c$, which is enhanced by contributions proportional to $\chi^s \chi^s$, as discussed in Appendix~\ref{app-BS}. 
        This paramagnon interference mechanism naturally induces BO fluctuations.

    \subsection{Numerical results for vertex functions}
        \begin{figure*}[!tb]
            \includegraphics[width=0.8\linewidth]{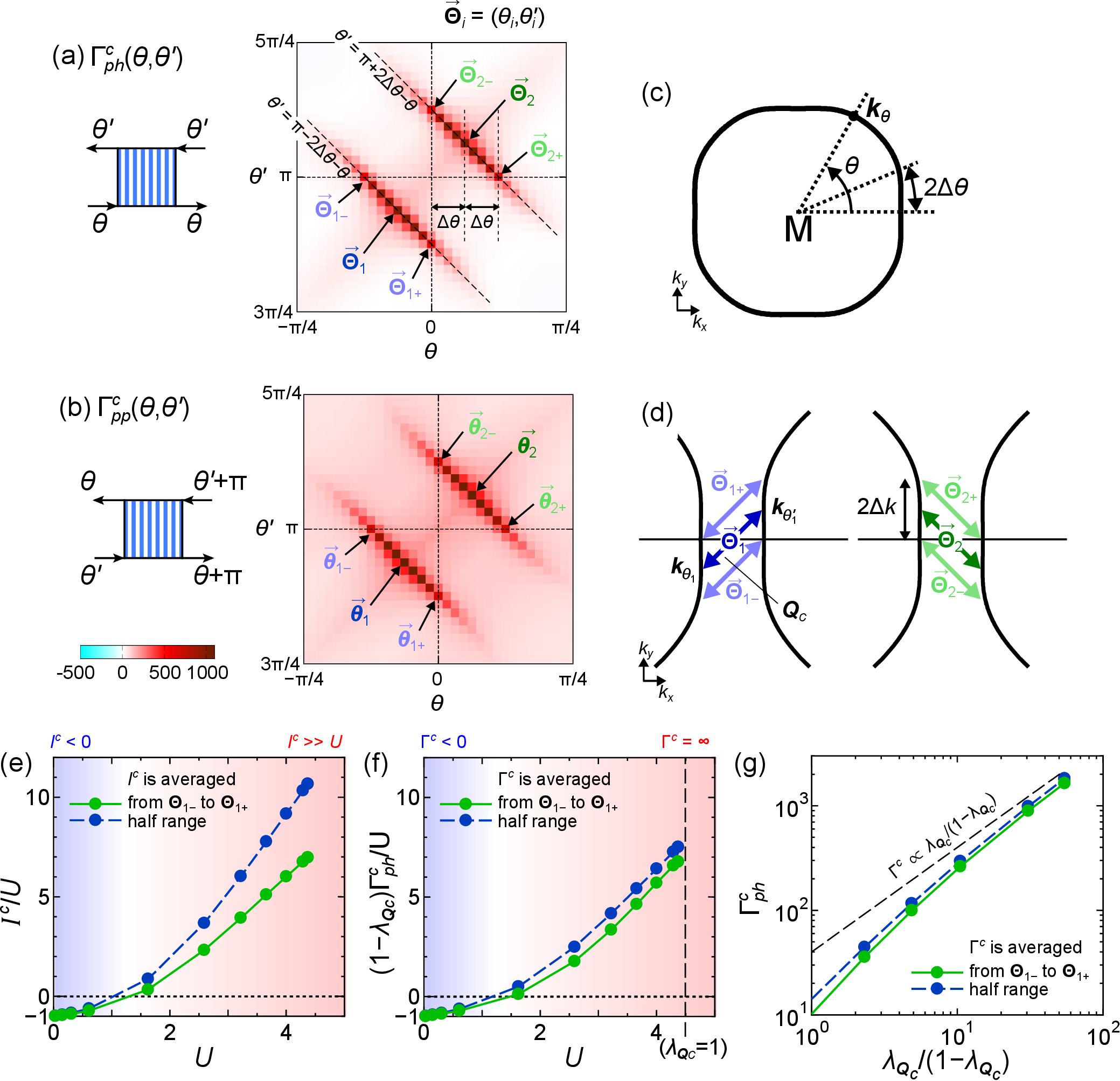}
            \caption{
                (a) Particle-hole (p-h) channel vertex $\Gamma_{ph}^c(\theta,\theta')$ and (b) particle-particle (p-p) channel vertex $\Gamma_{pp}^c(\theta,\theta')$ obtained from the BS equation for $U=4.29$, $T=0.01$, and $p=0.2$. 
                Both $\Gamma_{ph}^c$ and $\Gamma_{pp}^c$ are strongly enhanced and positive near the points $\bm{\Theta}_{1,2} = (\theta_{1,2},\theta_{1,2}')$, 
                where $\theta$ denotes the FS angle shown in (c) and $\epsilon_n=\epsilon_{n'}=\pi T$. 
                (c) Fermi surface and definition of $\theta$. 
                (d) Schematic illustration of the nesting condition: 
                between $\Theta_{1-}$ and $\Theta_{1+}$ and between $\Theta_{2-}$ and $\Theta_{2+}$, 
                the relation $\bm{k}_\theta - \bm{k}_{\theta'} \approx \bm{Q}_c$ is satisfied, 
                giving strong $\Gamma$ over a range $\pm 2\Delta k \sim \pm 2 k_F\Delta\theta$. 
                (e) $U$ dependence of the normalized irreducible interaction $I^c/U$ averaged along the dashed line in (a), 
                where $I^c = \frac12(I_{ph}^c + I_{pp}^c)$. 
                (f) $U$ dependence of the normalized full vertex $(1-\lambda_{\bm{Q}_c})\Gamma_{ph}^c/U$, 
                consistent with $\Gamma^c \approx I^c/(1-\lambda_{\bm{Q}_c})$. 
                The same holds for $\Gamma_{pp}^c$. 
                (g) $\Gamma_{ph}^c$ as a function of $\lambda_{\bm{Q}_c}/(1-\lambda_{\bm{Q}_c})$, 
                showing proportionality for $\lambda_{\bm{Q}_c} \lesssim 1$.
            }\label{fig3}
        \end{figure*}

        From Eq.~\eqref{eq-gamma0}, the four-point vertex 
        $\Gamma^x(k_1,k_2;k_3,k_4) = \Gamma^x_{k_3-k_4}(k_2,k_4)\,\delta_{k_1-k_2,k_3-k_4}$
        can be obtained for $x=c,s$.  
        Depending on the physical quantity of interest, different momentum combinations are needed.  
        Here we introduce the p-h and p-p channel vertices:
        \begin{align}
            \Gamma_{ph}^x(k,k') &\equiv \Gamma^x(k',k;k',k), \\
            \Gamma_{pp}^x(k,k') &\equiv \Gamma^x(k,k';-k',-k).
        \end{align}
        These interactions naturally incorporate CDW fluctuations arising from the AL term and can therefore describe CDW critical phenomena.

        Figures~\ref{fig3}(a) and \ref{fig3}(b) show $\Gamma_{ph}^c(\theta,\theta')$ and $\Gamma_{pp}^c(\theta,\theta')$ obtained from the BS equation for $U=4.29$, $T=0.01$, and $p=0.2$, with $\bm{k}$ and $\bm{k}'$ restricted to the Fermi surface [Fig.~\ref{fig3}(c)] and $\epsilon_n=\epsilon_{n'}=\pi T$.  
        In both channels, $\Gamma^c$ is strongly enhanced for $(\theta,\theta')$ satisfying $\bm{k}_\theta - \bm{k}_{\theta'} \approx \bm{Q}_c$, in agreement with Eq.~\eqref{eq-gamma-approx}.  
        This indicates the development of an effective interaction mediated by BO fluctuations.  
        The nesting geometry is illustrated in Fig.~\ref{fig3}(d), which will be shown in the next section to account for the enhancement of the effective mass and $d$-wave pairing near the CDW QCP.

        Figure~\ref{fig3}(e) shows the $U$ dependence of $I^c=(I_{ph}^c+I_{pp}^c)/2$ averaged along the dashed line in Fig.~\ref{fig3}(a).  
        In the weak-coupling regime $U\ll 1$, the RPA relation $I^c=-U$ holds.  
        As $U$ increases, vertex corrections, especially the AL terms, make $I^c$ positive and even $I^c \gg U$~\cite{Tazai2022_kagome,Tazai2016_fRG_orbital,Tazai2017_UVC,Tazai2019_multipole,Yamakawa2017_FeSe_SC,Onari2014_LaFeAsOH}.  
        Consequently, $\Gamma^c$ given by Eq.~\eqref{eq-gamma-approx} becomes crucial near the CDW QCP.

        The $U$ dependence of the normalized full vertex 
        $(1-\lambda_{\bm{Q}_c})\Gamma_{ph}^c/U$ is plotted in Fig.~\ref{fig3}(f),  
        showing a close similarity to $I^c/U$.  
        Figure~\ref{fig3}(g) further demonstrates that $\Gamma^c$ scales as  
        $\lambda_{\bm{Q}_c}/(1-\lambda_{\bm{Q}_c})$ for $\lambda_{\bm{Q}_c}\lesssim 1$,  
        confirming the validity of Eq.~\eqref{eq-gamma-approx}.  

        We note that $\Gamma^c$ obtained from the BS equation is strongly enhanced near $\lambda_{\bm{Q}_c}\approx 1$ primarily
        in the low-frequency region. 
        Therefore, in practice we employ a lowest-Matsubara-frequency approximation, 
        where $\Gamma^c$ at $|\epsilon_n|=\pi T$ is computed from the BS equation including vertex corrections, while other components are taken from FLEX. 
        Namely, the approximation is applied only to the external-frequency dependence used in subsequent calculations, not to the internal summations in the BS equation itself. 
        This approximation is justified near the QCP, where the characteristic energy scale of CDW fluctuations is low,
        and it provides a conservative estimate of vertex-correction effects, since the contribution of CDW fluctuations is expected to be underestimated within this approximation.

\section{Mass enhancement}\label{sec4}
    \begin{figure}[!htb]
        \includegraphics[width=0.9\linewidth]{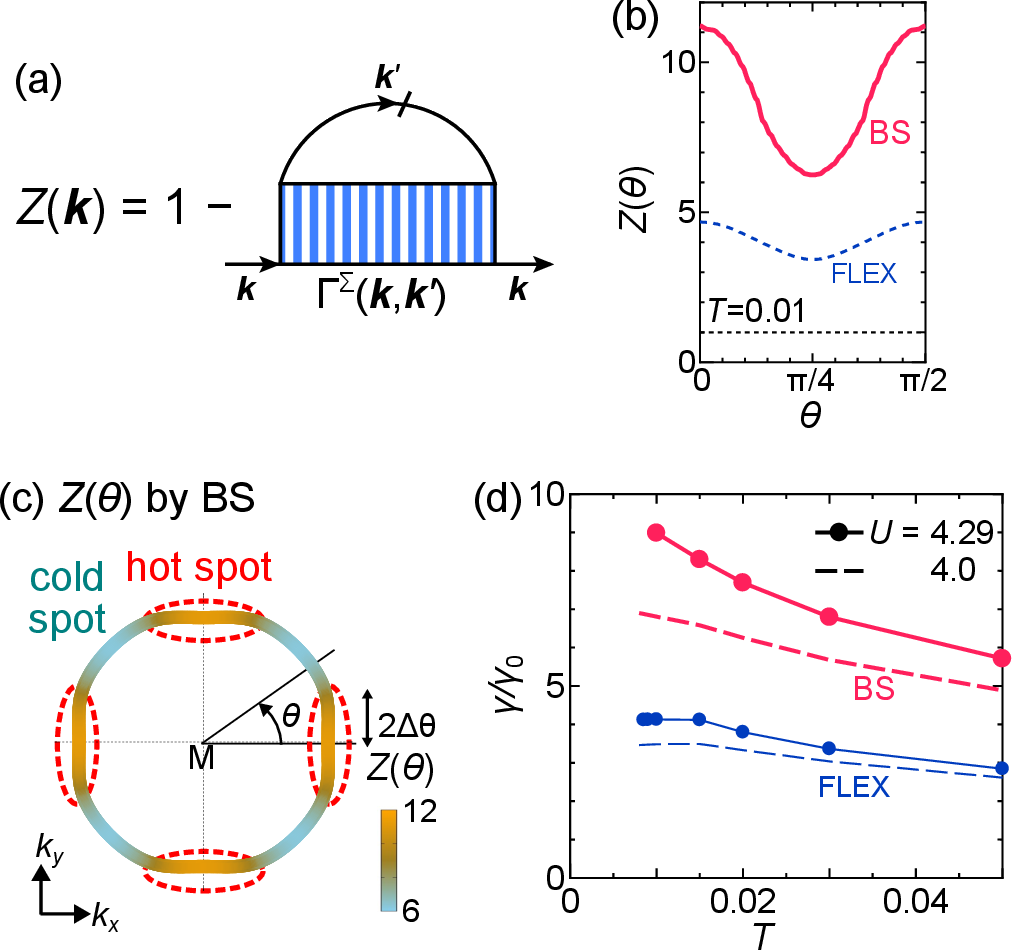}
        \caption{
            (a) Diagram of the mass-enhancement factor $Z (\bm{k})$ using $\Gamma^{\Sigma}$. 
            The solid line with a slash represents $\frac{\partial G(k)}{i \partial \omega}$. 
            The precise definition of $\Gamma^{\Sigma}$ is given in Appendix \ref{app-Z}. 
            (b) $Z(\theta)$ obtained by BS (red solid line) and $Z_{\mathrm{FLEX}} (\theta)$ obtained by FLEX (blue dashed line) for $p = 0.2$, $U = 4.29$, and various $T$. 
            (c) Color plot of $Z(\theta)$ obtained by BS. 
            Due to the strong peak in $\Gamma_{ph}^c$ originating from BO
            fluctuations, $Z(\theta)$ is significantly enhanced within the angular range $2 \Delta \theta$. 
            (d) Temperature dependence of the enhancement of the electronic specific-heat coefficient $\gamma / \gamma_0$ obtained by BS for $U=4.0$ and $4.29$ with $p = 0.2$. 
            The result $\gamma_{\mathrm{FLEX}} / \gamma_0$ obtained by FLEX is also shown. 
        }\label{fig4}
    \end{figure}

    Using the $\Gamma^c$ obtained by solving the BS equation in the previous section, we now evaluate the enhancement of the quasiparticle effective mass.  
    In strongly correlated electron systems, the effective mass $m^*$ is significantly increased, strongly influencing many physical properties such as the specific heat and superconductivity.  

    Figure~\ref{fig4}(a) shows the Ward identity for the mass-enhancement factor $Z = m^* / m$:  
    \begin{align}
        Z(\bm{k}) = 1 - \left. \frac{\partial}{i \partial \omega} \Sigma (\bm{k}, i \omega) \right|_{\omega = 0}.
    \end{align}
    Here, the solid line with a slash represents $\left. \frac{\partial}{i \partial \omega} G (\bm{k}, i \omega) \right|_{\omega = 0}$.  
    The vertex 
    $\Gamma^\mathrm{\Sigma}$
    corresponds to the four-point vertex given in Eq.~\eqref{eq-gamma0}, which includes interactions beyond the Migdal approximation (FLEX).  
    A more detailed explanation is provided in Appendix \ref{app-Z}.  

    The angular dependence of $Z(\theta)$ along the Fermi surface for $U=4.29$, $T=0.01$, and $p=0.2$ is shown in Fig.~\ref{fig4}(b).  
    Here we compare the results obtained by BS with those obtained by FLEX ($Z_{\mathrm{FLEX}}$).  
    When the CDW fluctuations are weak, $Z(\theta) \gtrsim Z_{\rm FLEX}(\theta)$.  
    In contrast, when the CDW fluctuations are strong, $Z(\theta) \gg Z_{\rm FLEX}(\theta)$ around the hot spots.  
    Figure~\ref{fig4}(c) presents a color plot of $Z(\theta)$ obtained by the BS equation.  
    In the BS-eq. method, the remarkable enhancement of $\Gamma_{ph}^c(\theta,\theta')$ around $\bm{k}-\bm{k'} \sim \bm{Q}_c$ [see Fig.~\ref{fig3}(a)] leads to strong mass enhancement at the hot spots.  
    In FLEX, by contrast, the angular dependence of $Z(\theta)$ remains relatively weak \cite{Bickers1991_FLEX,Kontani1999_Hall}.  

    Furthermore, Fig.~\ref{fig4}(d) shows the temperature dependence of the electronic specific-heat coefficient, $\gamma = \frac{1}{N}\sum_{\bm{k}} Z_{\bm{k}} \rho_{\bm{k}}$, where $\rho_{\bm{k}} = \delta(\mu - \epsilon_{\bm{k}})$. 
    Note that the specific heat coefficient of the noninteracting system is given by $\gamma_0 = (1/N) \sum_{\bm{k}} \rho_{\bm{k}}$.
    The BS result shows that $\gamma$ is enhanced compared to FLEX due to charge-channel fluctuations.  

    Near the CDW-QCP, at low temperatures ($T < 0.01$), quantum critical effects beyond the lowest-Matsubara-frequency approximation are expected to give rise to a $\log T$ dependence \cite{Moriya2000_highTc, Moriya1985}.  
    This issue is left as a future problem.

\section{Superconductivity}\label{sec5}
    \begin{figure}[!htb]
        \includegraphics[width=1\linewidth]{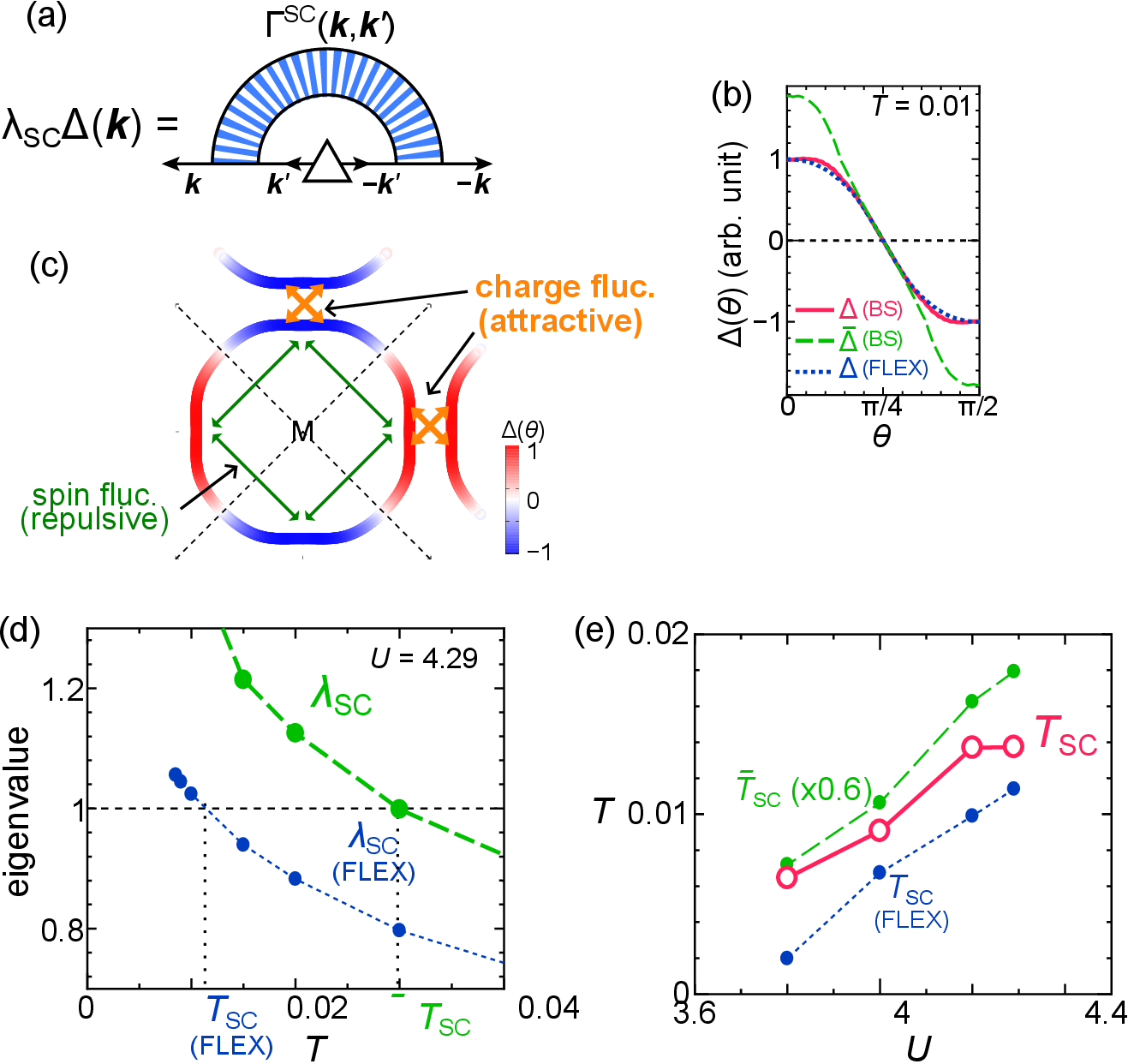}
        \caption{
            (a) Diagram of the linearized gap equation using $\Gamma$. 
            (b) Angular dependence of the gap functions $\Delta$ and $\bar{\Delta}$ by BS for $U = 4.29$, $T = 0.01$, and $p = 0.2$, where the renormalized gap function is $\Delta (\theta) = \bar\Delta (\theta)/Z (\theta)$. 
            The result of FLEX is also shown. 
            The magnitude of $\Delta$ is normalized by the slope at $\theta = \pi/4$.
            (c) $d$-wave superconducting structure mediated by spin and charge fluctuations. 
            (d) $T$-dependence of $\lambda_{\mathrm{SC}}$ obtained by BS and FLEX. 
            (e) $U$-dependence of $T_{\mathrm{SC}}$ and $\bar{T}_{\mathrm{SC}}$ obtained by BS, where $T_{\mathrm{SC}} = \bar{T}_{\mathrm{SC}} \cdot (\gamma_{\mathrm{FLEX}}/\gamma)$. 
            $T_{\mathrm{SC}}$ obtained by FLEX is also shown. 
        }\label{fig5}
    \end{figure}

    In this section, we analyze the superconducting state by solving the linearized gap equation using the interaction $\Gamma$ obtained from the BS equation.  

    In the conventional Migdal approximation, the electron-fluctuation coupling constant $U$ is fixed, and only the mediating effect of fluctuations is taken into account.  
    In contrast, we previously introduced the $U$-VC theory, in which the vertex correction $\Lambda^c$ at the nodal point is included beyond Migdal.  
    There, it was shown that the coupling constant between charge-channel fluctuations and Cooper pairs is significantly enhanced to $U \Lambda^c$ ($\Lambda^c \gg 1$) due to the AL-type vertex correction (AL-VC) \cite{Yamakawa2020_FeSe,Yamakawa2017_FeSe_P, Yamakawa2017_FeSe_SC, Tazai2017_UVC, Tazai2016_fRG_orbital, Onari2014_LaFeAsOH}.  
    In the present BS-eq. method, as shown in Fig.~\ref{fig3}(d), the irreducible vertex $I^c$ is strongly enhanced by the AL-type vertex correction, leading to the spontaneous formation of strong coupling between charge fluctuations and electrons.  
    Thus, the BS framework naturally extends the $U$-VC theory by including its diagrams.  

    The diagrammatic representation of the gap equation using $\Gamma$ is shown in Fig.~\ref{fig5}(a).
    Superconductivity emerges when the eigenvalue $\lambda_{\mathrm{SC}}$ reaches unity.  
    Here, the self-energy from FLEX is used for the Green's function, and the lowest-Matsubara-frequency approximation is applied to $\Gamma$ (see Appendix \ref{app-SC}).  

    The angular dependence of $\Delta(\theta)$ is shown in Fig.~\ref{fig5}(b).
    The renormalized gap function, $\Delta(\theta) = \frac{\bar{\Delta}(\theta)}{Z(\theta)}$, shows that the attractive contribution from $\Gamma^c$ is partially suppressed by mass renormalization, resulting in a $d$-wave superconducting gap similar to that obtained by FLEX (with the Migdal approximation). 

    Figure~\ref{fig5}(c) illustrates that spin and charge fluctuations act at different wave vectors and cooperatively stabilize $d$-wave superconductivity.  
    While the spin channel provides repulsion near $\bm{Q}_s$, the charge channel supplies attraction near $\bm{Q}_c$, 
    naturally explaining the strong enhancement of $d$-wave superconductivity near the CDW-QCP. 

    The $U$-dependence of the eigenvalue $\lambda_{\mathrm{SC}}$ is presented in Fig.~\ref{fig5}(d).
    Compared with the spin-fluctuation theory based on FLEX, the BS framework shows a pronounced enhancement of $\lambda_{\mathrm{SC}}$ and the transition temperature $\bar{T}_{\mathrm{SC}}$ due to the bond-order fluctuation interaction $V_{\rm BO}$.  
    Importantly, the attractive interaction from BO fluctuations is comparable in magnitude to the repulsive interaction from spin fluctuations, and both channels cooperate to realize high-$T_c$ $d$-wave superconductivity.  

    According to Fermi-liquid theory, energy scales such as the bandwidth are renormalized by a factor of $Z^{-1}$.  
    Consequently, the superconducting transition temperature is also renormalized by $Z^{-1}$.  
    In the present calculation, the Green's functions $G(\pm p)$ in the gap equation [Eq.~\eqref{eq-gap}] include the factor $Z_{\mathrm{FLEX}}^{-1}$.  
    However, once the BO-fluctuation contribution is included, the relevant renormalization factor becomes $Z_{\mathrm{BS}}^{-1}$.  
    Thus, the superconducting transition temperature corrected for BO fluctuations is given by
    \begin{align} \label{eq-tsc}
        T_{\mathrm{SC}} = \bar{T}_{\mathrm{SC}} \cdot \frac{Z_{\mathrm{FLEX}}}{Z_{\mathrm{BS}}}, 
    \end{align}
    where $\frac{Z_{\mathrm{FLEX}}}{Z_{\mathrm{BS}}} = \frac{\gamma_{\mathrm{FLEX}}}{\gamma_{\mathrm{BS}}}$. 
    Figure~\ref{fig5}(e) shows the $U$ dependence of $\bar{T}_{\mathrm{SC}}$, $T_{\mathrm{SC}}$, and $T^{\mathrm{FLEX}}_{\mathrm{SC}}$ from Migdal theory.  
    Here, $\gamma$ is evaluated at $T = 0.01$.  
    Although $\bar{T}_{\mathrm{SC}} \sim 3 T^{\mathrm{FLEX}}_{\mathrm{SC}}$, the inclusion of BO-induced renormalization reduces it to $T_{\mathrm{SC}} \gtrsim T^{\mathrm{FLEX}}_{\mathrm{SC}}$.  

    In summary, high-$T_c$ $d$-wave superconductivity emerges from the cooperation of spin and charge channel fluctuations near CDW-QCP.

\section{Phase diagram}\label{sec6}

    \begin{figure}[!htb]
        \includegraphics[width=1\linewidth]{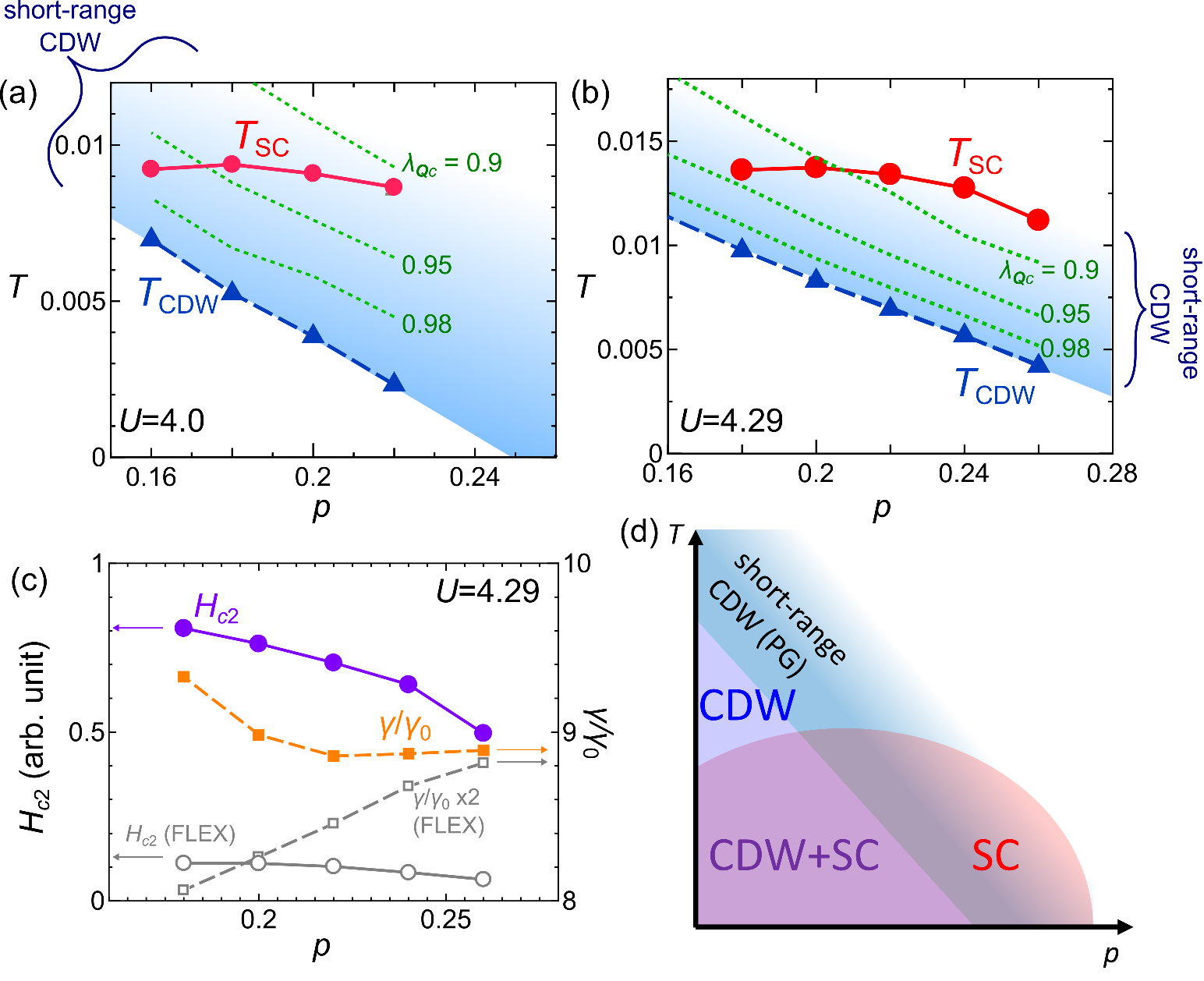}
        \caption{
            (a)(b) Doping dependence of the transition temperatures obtained from the present theory for (a) $U=4$ and (b) $U=4.29$.
            The blue dashed line with triangles represents $T_{\mathrm{CDW}}$ determined from $\lambda_{\bm{Q}_c}=1$,
            while the red solid line with circles denotes $T_{\mathrm{SC}}$.
            The green dashed lines correspond to $\lambda_{\bm{Q}_c}=0.98, 0.95, 0.9$.
            (c) Doping dependence of $H_{c2} \propto (T_c \gamma/\gamma_0 )^2$ and $\gamma/\gamma_0$ at $U=4.29$.
            Within the 
            BS-eq. method,
            both $H_{c2}$ and $\gamma$ are strongly enhanced compared to those obtained by FLEX, owing to the development of BO fluctuations.
            (d) Schematic phase diagram, where $T_{\mathrm{SC}}$ is enhanced in the vicinity of the CDW phase.
        }\label{fig6}
    \end{figure}

    Figure~\ref{fig6}(a) shows the phase diagram obtained from the BS equation for $U=4$.
    We plot the $p$-dependence of $T_{\mathrm{CDW}}$ obtained from the DW equation and $T_{\mathrm{SC}}$ obtained from Eq.~(\ref{eq-tsc}).
    Here, $T_{\mathrm{CDW}}$ is a monotonically decreasing function of $p$, whereas $T_{\mathrm{SC}}$ exhibits a gentle peak around $p \sim 0.18$.
    Moreover, $T_{\mathrm{CDW}} \sim T_{\mathrm{SC}}$ at $p \sim 0.15$, and $T_{\mathrm{CDW}} < T_{\mathrm{SC}}$ for $p > 0.15$.
    These results are consistent with experiments~\cite{Keimer2015_Review}.
    In Fig.~\ref{fig6}(a), in addition to $T_{\mathrm{CDW}}$ ($\lambda = 1$), we also show the $p$-dependence for $\lambda = 0.98, 0.95,$ and $0.9$.
    In regions where CDW fluctuations are well developed, nanoscale CDW domains may emerge around a small amount of impurities and can give rise to a pseudogap.
    Figure~\ref{fig6}(b) shows the phase diagram for $U=4.29$, where $T_{\mathrm{SC}}$ is enhanced.
    The obtained value of $T_{\mathrm{SC}} \sim 0.01 |t|$ ($\sim 40$~K) corresponds to that of the single-layer cuprate LSCO.
    It should be noted that the competition between superconductivity and CDW leads to the suppression of $T_{\mathrm{CDW}}$ within the superconducting phase and vice versa.

    The doping dependence of the orbital upper critical field $H_{c2}$ is shown in Figure~\ref{fig6}(c). 
    It is estimated from $H_{c2} = \phi_0 / (2\pi \xi_{\mathrm{GL}}^2)$ with 
    $\xi_{\mathrm{GL}} = \hbar v_F / (\pi \Delta_0)$, 
    giving $H_{c2} \propto (T_c \gamma / \gamma_0)^2$. 
    Here, $\gamma$ at $T=0.01$ is used. 
    Within the BS-eq. method including charge-channel interactions, 
    both $H_{c2}$ and $\gamma$ are strongly enhanced compared with FLEX.

    A schematic phase diagram is shown in Fig.~\ref{fig6}(d).
    Near the CDW quantum critical point, a pseudogap region emerges due to short-range CDW correlations, and a $d$-wave high-$T_c$ superconducting phase is realized through the cooperative effects of BO and spin fluctuations.

\section{Discussion and Conclusions}\label{sec7}
    We have developed the BS-eq. method that enables a quantitative evaluation of the effective interaction $V_{\mathrm{BO}}$ beyond the mean-field approximation.
    This theory, based on the Baym-Kadanoff conserving approximation, systematically includes vertex corrections such as the Maki-Thompson and Aslamazov-Larkin terms.
    As a result, it provides a microscopic and systematic framework to derive interactions mediated by fluctuations,
    without introducing adjustable parameters other than $U$.
    The calculated $V_{\mathrm{BO}}$ shows strong momentum dependence and exhibits a divergent enhancement near the CDW-QCP.
    The enhancement of $V_{\mathrm{BO}}$ by vertex corrections agrees with previous theoretical studies \cite{Tazai2017_UVC,Tazai2018_fulls,Tazai2019_multipole,Tazai2022_kagome}, supporting the reliability of the present framework.

    By applying this method to the single-band Hubbard model, we showed that spin and BO fluctuations cooperate to strongly enhance $d$-wave superconductivity near the BO QCP ($p \sim 0.2$).
    In this region, both the superconducting transition temperature $T_c$ and the upper critical field $H_{c2}$ are strongly enhanced, accompanied by a large increase in the effective mass $m^*$ ($Z$).
    These results provide a consistent microscopic picture of superconductivity and quantum criticality in cuprates. 
    The possible effects of bilayer Fermi-surface splitting \cite{Hossain2008ARPES} would be an interesting future issue.

    The BS approach can also be applied to multiorbital systems. 
    Indeed, it has recently been used to study nickelates, 
    where a remarkable result has been obtained: an $s_{\pm}$-wave high-$T_c$ superconducting mechanism driven by the cooperation of CDW and SDW fluctuations~\cite{Inoue2026_Ni_CDW}. 
    Its application to iron-based superconductors, where nematic and smectic fluctuations are important, is a promising future direction.

    In dynamical mean-field theory, extensions beyond the local approximation have been developed~\cite{Maier2005_cDMFT,Schafer2021_cDMFT}, such as the Dynamical Cluster Approximation (DCA)~\cite{Maier2000_DCA,Jarrell2001_DCA}, Cellular DMFT~\cite{Kotliar2001_cDMFT,Sordi2012_cDMFT,Civelli2005_cDMFT}, and the Dynamical Vertex Approximation (D$\Gamma$A)~\cite{Toschi2007_DGA,Katanin2009_DGA,Rohringer2018_DGA}.
    In these methods, a dome-shaped $d$-wave superconducting phase has been obtained due to spin fluctuations. On the other hand, in the present BS-eq. method, near optimal doping, we found (1) the emergence of nonmagnetic CDW and (2) the enhancement of the transition temperature of $d$-wave superconductivity due to CDW fluctuations. This is consistent with experimental results showing that $T_c$ becomes maximal near the QCP of the CDW in cuprates.

    The present BS-eq. method is based on the Luttinger-Ward functional derived from the FLEX self-energy, and is therefore applicable from the weak- to intermediate-coupling regimes, targeting the optimal-to-overdoped regions in cuprates. 
    On the other hand, the two-particle self-consistent (TPSC) theory is a method that can describe spin fluctuations near half filling~\cite{Vilk1997_TPSC,Otsuki2012_TPSC,Simard2023_TPSC}. 
    The combination of TPSC and the present BS-eq. method may provide a route to describing CDW while extending applicability to the strong-coupling regime.

    In the present numerical calculations, we first obtain $\Gamma_q(k,k')$ by numerically solving the BS-eq. \eqref{eq-gamma0}, where the internal summations are performed exactly. 
    Then, the lowest Matsubara-frequency approximation is applied only to the external-frequency dependence of $\Gamma^\mathrm{SC}(k,p)$ and $\Gamma^\mathrm{\Sigma}(k,p)$, and the superconducting gap equation and the self-energy are solved. 
    This approximation is based on the fact that quantum critical fluctuations and superconductivity are governed by low-energy processes, whereas contributions from the high-frequency region are non-singular. 
    Although this approximation underestimates the contribution of charge-channel fluctuations, their contribution to the pairing interaction is important. 
    An analysis including general frequencies significantly increases the computational cost and remains an important issue for future work.

\begin{acknowledgments}
    The authors are grateful to Rina Tazai and Seiichiro Onari for fruitful discussions. 
    This work was supported by JSPS KAKENHI (Grant Numbers JP20K03858, JP24K00568, and JP24K06938).
\end{acknowledgments}

\section*{DATA AVAILABILITY}

The data that support the findings of this article are not publicly available. 
The data are available from the authors upon reasonable request.


\appendix

\section{Formulation of the Bethe-Salpeter Equation}\label{app-BS}
\begin{figure*}[!htb]
        \includegraphics[width=0.8\linewidth]{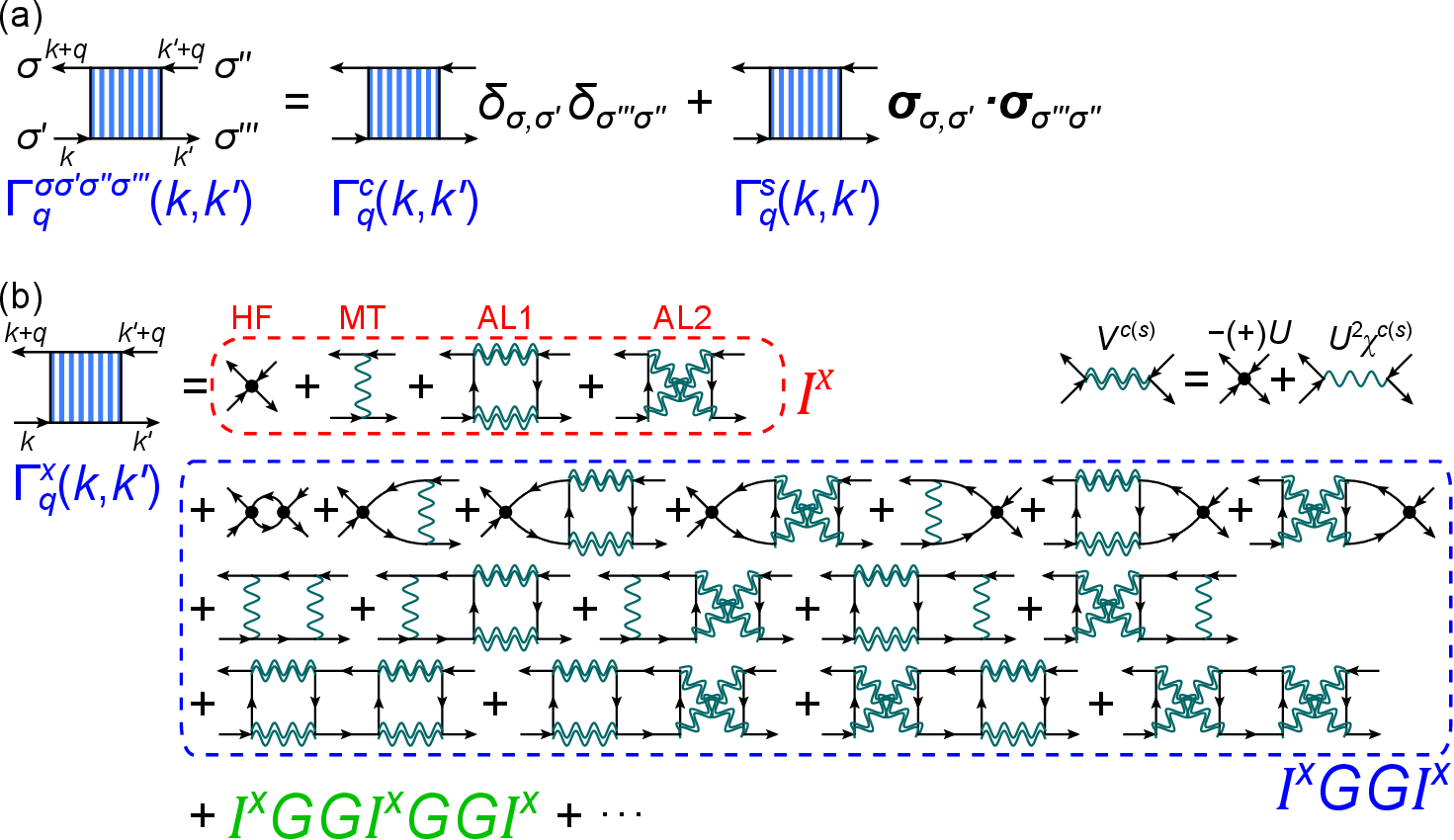}
        \caption{
        (a) The full four-point vertex $\Gamma^{\sigma \sigma' \sigma'' \sigma'''}_q (k,k')$. 
        Under SU(2) symmetry, it is represented by $\Gamma^c$ and $\Gamma^s$.
        (b) Diagrammatic expression of $\Gamma$ obtained by solving the BS equation.
        The double wavy line denotes $V^{c(s)} = -(+) U + U^2 \chi^{c(s)}$.
        }\label{fig7}
    \end{figure*}

    In this study, we introduce the Bethe-Salpeter (BS) equation, which plays a central role, and determine the four-point vertex $\Gamma$ representing the full interaction.
    The BS equation is given by the following self-consistent form:
    \begin{align}
        &\Gamma^{\sigma \sigma' \sigma'' \sigma'''}_q (k,k')
        = I^{\sigma \sigma' \sigma'' \sigma'''}_q (k,k') \nonumber \\
        &+ \frac{T}{N} \sum_{k''} \sum_{\rho \rho' \rho'' \rho'''} 
        I^{\sigma \sigma' \rho \rho'}_q (k,k'') 
        G^{\rho \rho''}(k''+q) G^{\rho'''\rho'}(k'') \nonumber \\
        & \times \Gamma^{\rho'' \rho''' \sigma'' \sigma'''}_q (k'',k)
        \label{eq-gamma-ssss}
    \end{align}
    Here, $k = (\bm{k}, i\epsilon_n)$ and $q = (\bm{q}, i\omega_l)$ denote combined notations for momentum and Matsubara frequencies for fermion $\epsilon_n = (2n+1) \pi T$ and boson $\omega_l = 2l \pi T$, respectively.
    $\sigma$, $\rho = \uparrow, \downarrow$ denote the spin indices at each vertex, as illustrated in Fig.~\ref{fig7}.
    $I$ is the irreducible vertex, whose explicit form will be given later.
    Assuming that the system satisfies SU(2) symmetry, the spin dependence of $\Gamma$ can be expressed in terms of the charge-channel vertex $\Gamma^c$ and the spin-channel vertex $\Gamma^s$ as
    \begin{align}
        \Gamma^{\sigma \sigma' \sigma'' \sigma'''}_q (k,k'') 
        &= \Gamma^c_q (k,k'') \delta_{\sigma, \sigma'}  \delta_{\sigma''', \sigma''} \nonumber \\
        &+ \Gamma^s_q (k,k'') \bm{\sigma}_{\sigma, \sigma'} \cdot \bm{\sigma}_{\sigma''', \sigma''}
        \label{eq-gamma-spin}
    \end{align}    
    where $\bm{\sigma}$ represents the Pauli matrices. 

    Under SU(2) symmetry, the Green's function is $G^{\sigma \sigma'}(k) = G (k) \delta_{\sigma \sigma'}$, and the BS equation is separated into the charge channel ($x=c$) and spin channel ($x=s$) as
    \begin{align}
        \Gamma^x_q(k,k') &= I^x_q(k,k') \nonumber \\
        & - \frac{T}{N} \sum_{k''} I^x_q(k,k'') G(k''+q) G(k'') \Gamma^x_q(k'',k') 
        \label{eq-gamma}
    \end{align}
    From Eq.~\eqref{eq-gamma-spin}, the charge (spin) channel vertices $X^{c(s)} = \Gamma^{c(s)}, I^{c(s)}$ are given by
    \begin{align}
        X^c &= X^{\uparrow \uparrow \uparrow \uparrow} + X^{\uparrow \uparrow \downarrow \downarrow} \\
        X^s &= X^{\uparrow \uparrow \uparrow \uparrow} - X^{\uparrow \uparrow \downarrow \downarrow}
        = X^{\uparrow \downarrow \uparrow \downarrow}
    \end{align}

    The irreducible vertex $I$ appearing in Eq.~\eqref{eq-gamma-ssss} is defined as the functional derivative of the Luttinger-Ward functional $\Phi^{\rm LW}$.
    For example, in the case of $q = 0$, it is expressed as
    \begin{align}
        I^{\sigma \sigma' \sigma'' \sigma'''}_0 (k,k') 
        = \frac{\delta^2 \Phi^{\rm LW}[G]}{\delta G^{\sigma' \sigma}_{k} \delta G^{\sigma'' \sigma'''}_{k'}}
        \label{eq-ward}
    \end{align}

    In this study, we adopt the one-loop approximation for $\Phi^{\rm LW}$, which decomposes into the following four terms:
    \begin{align}
        I^x = I^{\mathrm{HF},x} + I^{\mathrm{MT},x} + I^{\mathrm{AL1},x} + I^{\mathrm{AL2},x}
    \label{eq-irr}
    \end{align}
    The Hartree-Fock (HF) term (Hartree term in the single-orbital case) is simply
    \begin{align}   
        I^{\mathrm{HF},c}_q(k,k') &= I^{\mathrm{HF},c} = -U \\
        I^{\mathrm{HF},s}_q(k,k') &= I^{\mathrm{HF},s} = U
    \end{align}
    The Maki-Thompson (MT) term corresponds to the one-magnon exchange process and is given by the sum (difference) of longitudinal $V^{\uparrow \uparrow \uparrow \uparrow}$ and transverse $V^{\uparrow \downarrow \uparrow \downarrow}$ particle-hole contributions in the charge (spin) channel. 
    The MT terms are given by
    \begin{align}
        I^{\mathrm{MT},c}_q(k,k') &= -\frac{1}{2} V^c(k - k') - \frac{3}{2} V^s(k - k') \nonumber \\
        & + U + U^2 \chi^0(k - k'), \\
        I^{\mathrm{MT},s}_q(k,k') &= -\frac{1}{2} V^c(k - k') + \frac{1}{2} V^s(k - k') - U.
        \label{eq-irrmt}
    \end{align}
    Here, $V^{c(s)}(q) = \pm U + U^2 \chi^{c(s)}(q)$ denotes the effective interaction in the charge (spin) channel obtained within FLEX.
    The terms $U$ and $U^2 \chi^0$ serve as corrections for double counting with the HF term and the AL1 term, respectively.

    The Aslamazov-Larkin (AL) terms correspond to two-magnon processes describing interference of spin fluctuations:
    \begin{align}
        I^{\mathrm{AL1},x}_q (k,k') &= \frac{T}{N} \sum_{q'} W^{x,1}(q,q') G(k - q') G(k' - q')
        \label{eq-irral1} \\
        I^{\mathrm{AL2},x}_q (k,k') &= \frac{T}{N} \sum_{q'} W^{x,2}(q,q') G(k + q + q') G(k' - q')
        \label{eq-irral2}
    \end{align}
    The functions $W^{x,n} (q,q')$ represent convolution terms of the spin and charge channels, given by
    \begin{align}
        W^{c,n}(q,q') &= \frac{1}{2} V^c(q - q') V^c(q') + \frac{3}{2} V^s(q - q') V^s(q') \nonumber \\ 
        &- U^2
        \label{eq-irralw-c} \\
        W^{s,n}(q,q') &= \frac{1}{2} V^c(q - q') V^s(q') + \frac{1}{2} V^s(q - q') V^c(q') \nonumber \\
        & - (-1)^n V^s(q - q') V^s(q') 
        \label{eq-irralw-s}
    \end{align}
    Here again, the term $-U^2$ is a correction for double counting.

    The diagrammatic structures incorporated by solving the BS equation are shown in Fig.~\ref{fig7}(b). 
    In RPA and FLEX approximations, the irreducible vertex $I$ is approximated only by the Hartree term, which leads to $\Gamma^x = V^x$. 
    In contrast, the higher-order terms beyond $V^x$ originate from vertex corrections, and they are automatically incorporated within the BS framework.

    Since $\chi^s$ is strongly enhanced within FLEX, the most significant contributions in the charge channel term of Eq.~\eqref{eq-irralw-c} are the AL1 and AL2 terms. 
    These describe the interference effects of two paramagnons via $W^c \propto V^s V^s$, which strongly enhance charge fluctuations~\cite{Kontani2021_review}.

    On the other hand, in Eq.~\eqref{eq-irralw-s}, the $\pm V^s V^s$ terms included in $W^{s}$ are small for even-parity spin fluctuations. 
    For ordinary spin fluctuations, the AL1 and AL2 terms largely cancel each other. 
    The $V^c V^s$ terms are small, and hence vertex corrections in the spin channel are weak~\cite{Tazai2016_fRG_orbital, Yamakawa2017_FeSe_SC}.
    Nevertheless, the $V^s V^s$ terms in Eq.~\eqref{eq-irralw-s} may potentially give rise to odd-parity orders such as spin-loop currents~\cite{Kontani2021_SLC}.

    The present theory, when combined with the one-loop self-energy $\Sigma_{\mathrm{FLEX}}$ from FLEX, satisfies the Baym-Kadanoff conservation laws (particle number, energy, and momentum conservation).

\section{Mass enhancement factor using $\Gamma$}\label{app-Z}

\begin{figure*}[!tb]
        \includegraphics[width=0.7\linewidth]{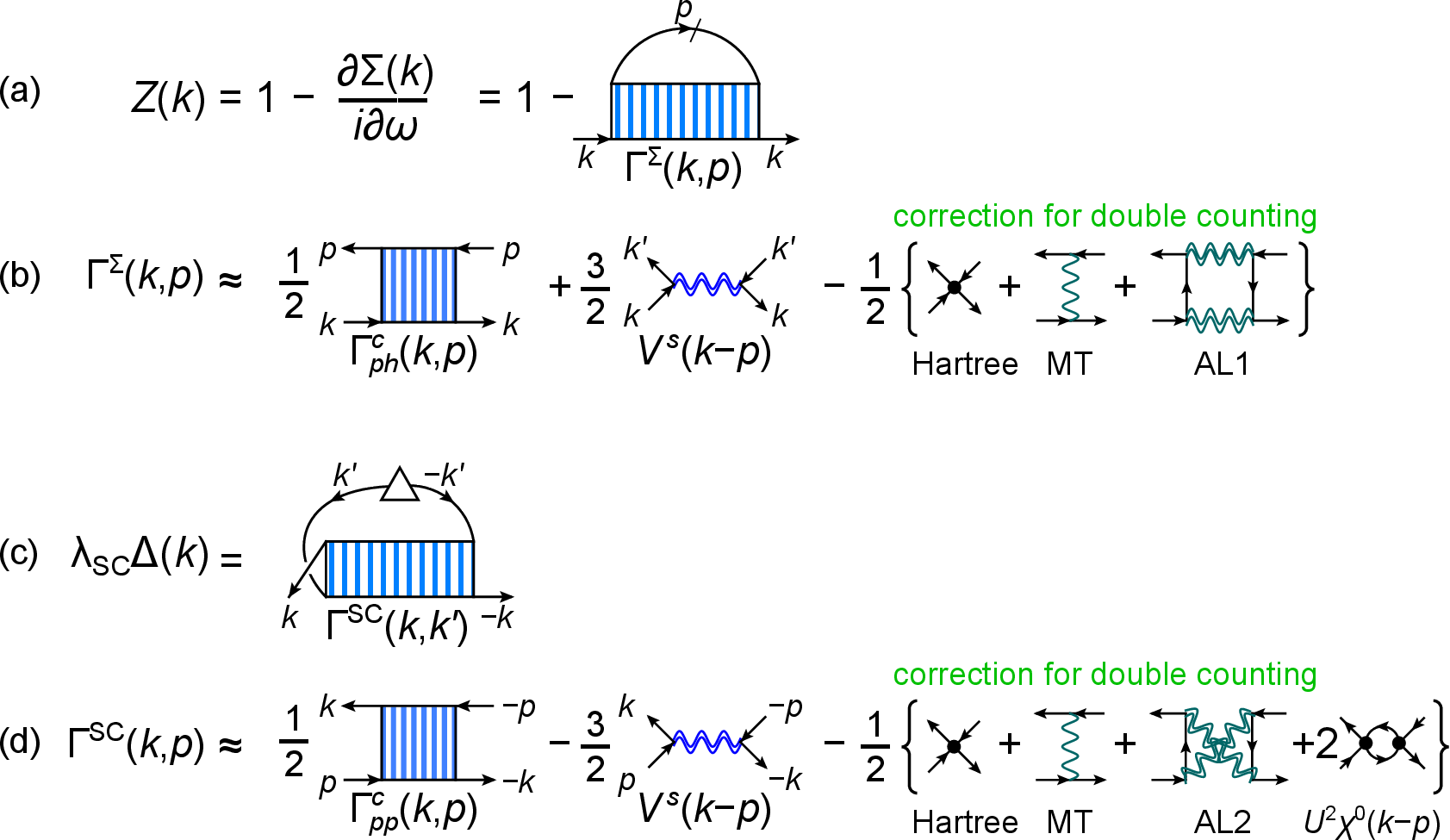}
        \caption{
            (a) Diagrammatic expression of $Z(k)$. 
            (b) Diagrammatic expression of $\Gamma^{\Sigma}(k,p)$ used in the present numerical calculation. 
            (c) Diagrammatic expression of the linearized gap equation. 
            (d) Diagrammatic expression of $\Gamma^{\mathrm{SC}}(k,p)$ used in the present numerical calculation. 
        }\label{fig8}
    \end{figure*}

    First, within Fermi liquid theory, the exact expression for the mass enhancement factor $Z$ is given by~\cite{Abrikosov1963}
    \begin{align}
        Z (k) = 1 - \frac{\partial \Sigma (k)}{i \partial \omega}
        = 1 - \frac{T}{N}\sum_{p} \tilde{I}^{ph} (k,p) \frac{\partial G (p)}{i \partial \omega},
    \end{align}
    where $\tilde{I}^{ph}$ denotes the exact particle-hole irreducible four-point vertex.
    The diagrammatic expression of $Z$ is shown in Fig.~\ref{fig8}(a). 

    Next, within the Bethe-Salpeter (BS) equation framework, $\tilde{I}^{ph}$ is approximated as
    \begin{align}
        \tilde{I}^{ph} (k,p) = \Gamma^{\Sigma} (k,p),
    \end{align}
    and
    \begin{align}
        \Gamma^{\Sigma} (k,p) = \tfrac{1}{2}\,\Gamma_{ph}^c(k,p) + \tfrac{3}{2}\,\Gamma_{ph}^s(k,p)
        - (\text{corrections}).
    \end{align}
    Here, $(\text{corrections})$ denotes terms that eliminate double-counting, 
    as shown in Fig.~\ref{fig8}(b).

    Finally, in the present numerical calculation, 
    \begin{align}
        Z (\bm{k}) = 1 - \frac{T}{N} \sum_{\bm{p}} \Gamma^{\Sigma} (\bm{k}, \bm{p}) \rho_{\bm{p}}.
    \end{align}
    Here, $\Gamma^{\Sigma} (\bm{k}, \bm{p})$ is the lowest-Matsubara-frequency approximation.
    Also, $\Gamma_{ph}^s$ is taken from FLEX (i.e., $\Gamma_{ph}^s = V^s$). 

    As an indicator of the effective mass enhancement, we also consider the electronic specific heat coefficient $\gamma$, which is given by the expectation value of $Z (\bm{k})$ on the Fermi surface:
    \begin{align}
        \gamma = \frac{1}{N} \sum_k Z (\bm{k}) \rho_{\bm{k}}.
    \end{align}
    Defining $\gamma_0$ as the specific heat coefficient without renormalization, the momentum-averaged mass enhancement factor is expressed as $Z \equiv \gamma / \gamma_0$.

\section{Linearized gap equation using $\Gamma$}\label{app-SC}

    First, within Fermi liquid theory, the exact expression for the linearized gap equation is \cite{Abrikosov1963}
    \begin{align} \label{eq-gap}
        \lambda_{\mathrm{SC}} \bar{\Delta} (k) = \frac{T}{N} \sum_{p} \tilde{I}^{pp} (k,p) G(p) \bar{\Delta} (p) G(-p),
    \end{align}
    where $\tilde{I}^{pp}$ denotes the exact particle-particle irreducible four-point vertex.
    The diagrammatic expression is shown in Fig.~\ref{fig8}(c). 

    Next, within the Bethe-Salpeter (BS) equation framework, $\tilde{I}^{pp}$ is approximated as
    \begin{align}
        \tilde{I}^{pp} (k,p) = \Gamma^{\mathrm{SC}}(k,p),
    \end{align}
    and
    \begin{align}
        \Gamma^{\mathrm{SC}} (k,p) = \tfrac{1}{2} \Gamma_{pp}^c (k,p) - \tfrac{3}{2} \Gamma_{pp}^s (k,p)
        - (\text{corrections}),
    \end{align}
    where $(\text{corrections})$ denotes terms that eliminate double-counting.

    Finally, in the present numerical calculation,
    \begin{align}
        \lambda_{\mathrm{SC}} \bar{\Delta} (k) = \frac{T}{N} \sum_{p} \Gamma^{\mathrm{SC}} (k, p) G(p) \bar{\Delta}(p) G(-p).
    \end{align}
    Here, $\Gamma^{\mathrm{SC}}(k, p)$ is obtained by the lowest-Matsubara-frequency approximation: 
    $\Gamma^c_{pp}$ at $|\epsilon_n|=\pi T$ is computed from the BS equation including vertex corrections, while other frequency components are taken from FLEX. 
    Also, $\Gamma_{pp}^s$ is taken from FLEX (i.e., $\Gamma_{pp}^s = V^s$). 
    The correction terms are shown in Fig.~\ref{fig8}(d). 
    Numerically, we solve the linearized gap equation on the FSs.

\section{Momentum dependence of $\lambda_q$}\label{app-DW}
    \begin{figure}[!tb]
        \includegraphics[width=1\linewidth]{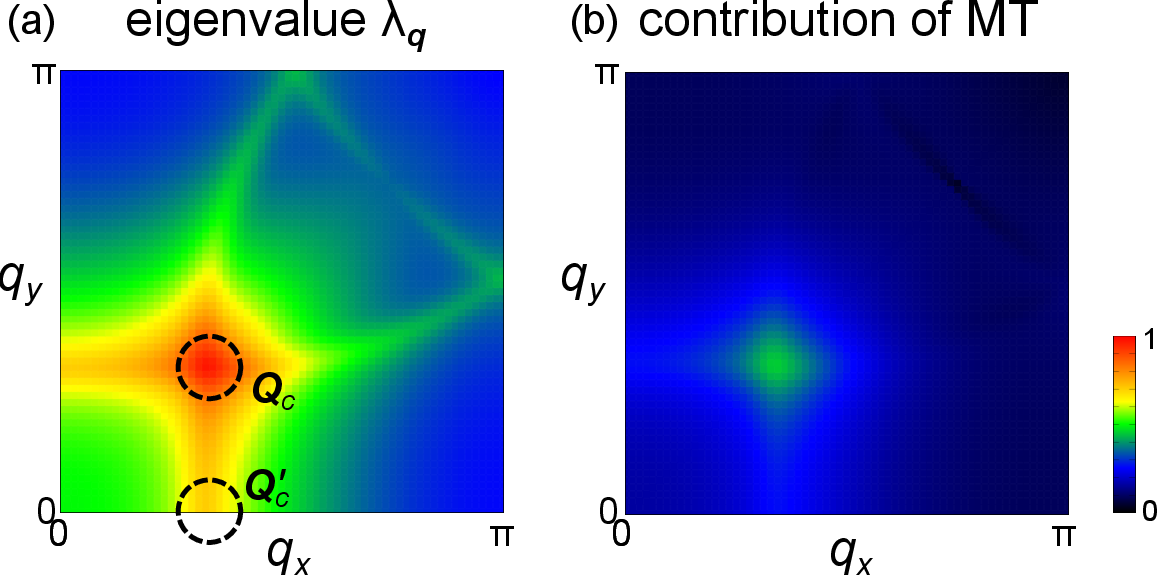}
        \caption{ 
            (a) $\bm{q}$-dependence of $\lambda_{\bm{q}}$. 
            (b) $\bm{q}$-dependence of the MT-term contribution to $\lambda_{\bm{q}}$. 
            Here, we set $U = 4.29$, $T = 0.01$, and $p = 0.2$. 
        }\label{fig9}
    \end{figure}

    In this Appendix, we present the full momentum dependence of the eigenvalue $\lambda_q$ of the DW equation~\eqref{eq-DW} and its MT-term contribution. 
    These results correspond to $\lambda_q$ shown in Fig.~\ref{fig2}(e) in the main text.

    Figure~\ref{fig9}(a) shows the two-dimensional $q$ dependence of $\lambda_q$ for $p=0.2$, $U=4.29$, and $T=0.01$. 
    $\lambda_q$ shows a broad enhancement from $\bm{Q}_c$ toward $\bm{Q}'_c$.

    Figure~\ref{fig9}(b) shows the $q$-dependence of the MT-term contribution to $\lambda_q$. 
    The MT contribution appears mainly near $\bm{Q}_c$ and is absent at $\bm{Q}'_c$, consistent with Ref.~\cite{Sachdev2013_CDW}.
    Its magnitude is also small. 
    Therefore, the dominant contribution to the charge-channel bond order originates from the AL terms~\cite{Yamakawa2015cuprate,Kawaguchi2017DW,Tsuchiizu2018cuprate}. 
    We note that the Hartree term hardly couples to the bond order, and its contribution is nearly zero (slightly negative).

\section{Supplementary analysis of charge-channel vertices}\label{app-I}
    \begin{figure}[!tb]
        \includegraphics[width=1\linewidth]{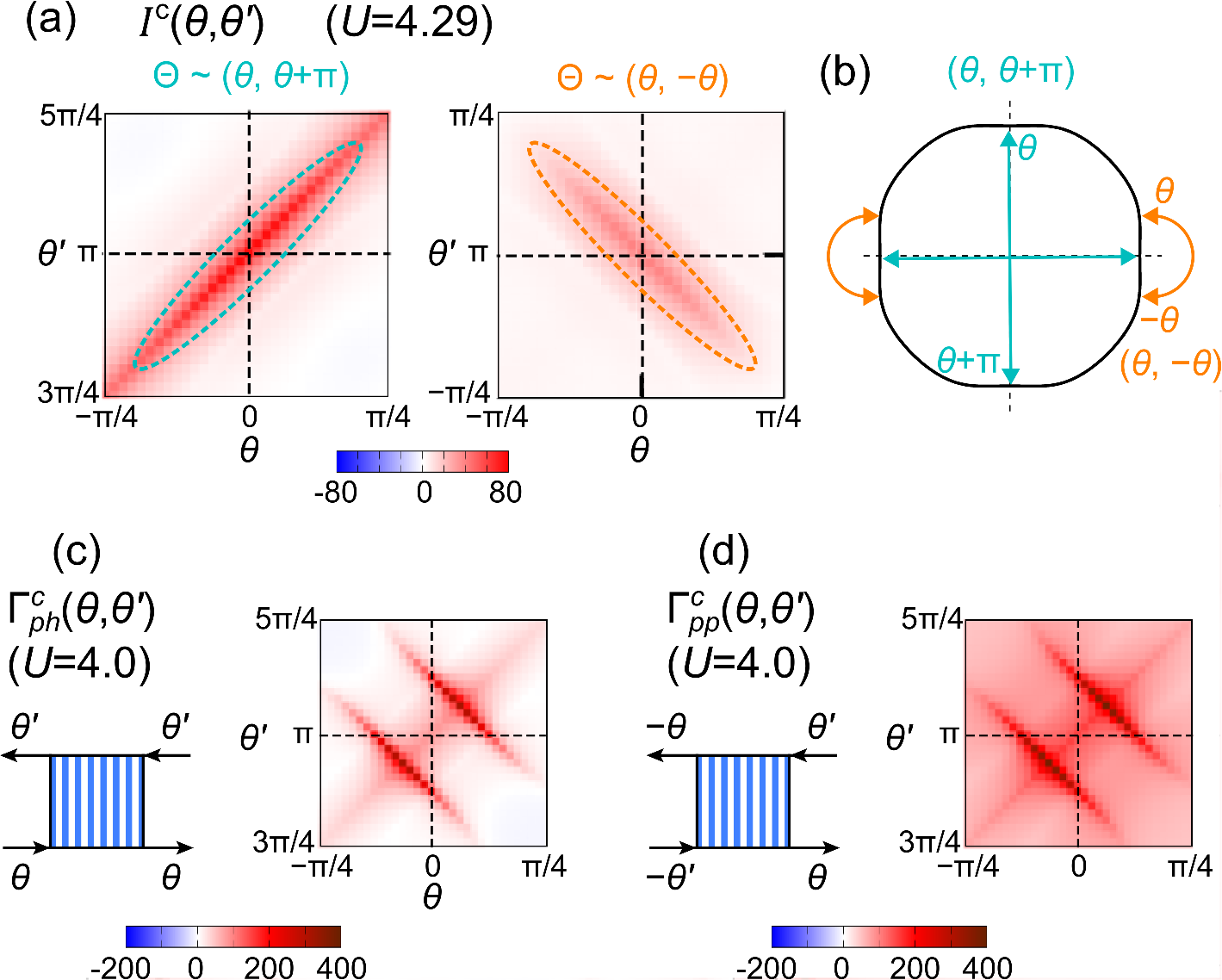}
        \caption{ 
            (a) $I^c = ( I^c_{ph} + I^c_{pp} ) /2$ for $U = 4.29$, $T = 0.01$, and $p = 0.2$.
            (b) Schematic illustration of the scattering process where $I^c$ develops.
            (c) $\Gamma_{ph}^{c}$ for $U = 4.0$.
            (d) $\Gamma_{pp}^{c}$ for $U = 4.0$.
        }\label{fig10}
    \end{figure}

    Figure~\ref{fig10}(a) shows $I^c (\theta, \theta')$ for $U=4.29$, $T=0.01$, and $p=0.2$.
    It takes large positive values along the lines satisfying $\theta = \theta' + \pi$ and $\theta = - \theta'$.
    Figure~\ref{fig10}(b) schematically illustrates the corresponding scattering process.
    This originates from the AL terms: 
    in Eqs.~\eqref{eq-irral1} and \eqref{eq-irral2}, 
    if the momentum dependence of $W$ is neglected, 
    the momentum summation of the product of internal Green's functions becomes large when $\bm{k} = - \bm{k}'$.
    For small $U$, one has $I^c \approx -U < 0$.
    Therefore, $I^c > 0$ is due to the vertex corrections.

    Figure~\ref{fig10}(c) shows $\Gamma^c_{\rm ph}$ for $U = 4.0$, $T = 0.01$, and $p = 0.2$.
    Figure~\ref{fig10}(d) shows $\Gamma^c_{\rm pp}$ for $U = 4.0$, $T = 0.01$, and $p = 0.2$.
    Although the absolute values are smaller than those at $U=4.29$ shown in Figs.~\ref{fig3}(a) and \ref{fig3}(b) in the main text, they exhibit very similar $\theta, \theta'$ dependence.

\bibliography{ref_cuprate, ref_various, ref_theory, ref_add}

\end{document}